\documentclass[conference]{ifacconf}
\usepackage{cite}
\usepackage{amsmath,amssymb,amsfonts}
\usepackage{graphicx}
\usepackage{natbib}\usepackage{savesym}\savesymbol{AND}
\usepackage{textcomp}
\usepackage{subfig}\usepackage{wrapfig}
\usepackage{mathrsfs}
\usepackage{algorithm}
\usepackage{algorithmic}
\def\BibTeX{{\rm B\kern-.05em{\sc i\kern-.025em b}\kern-.08em
    T\kern-.1667em\lower.7ex\hbox{E}\kern-.125emX}}
\newcommand{\bld}[1]{\boldsymbol{#1}}
\DeclareMathOperator*{\argmin}{argmin}

\begin{document}
	\begin{frontmatter}
\title{Estimating States and Model Uncertainties Jointly by a Sparsity Promoting UKF}

\author{Ricarda-Samantha Götte, Julia Timmermann} 

\address{Heinz Nixdorf Institute, Paderborn University, Paderborn, Germany
	(e-mail: \{rgoette,julia.timmermann\}@hni.upb.de)}

\begin{abstract}
State estimation when only a partial model of a considered system is available remains a major challenge in many engineering fields. This work proposes a joint, square-root unscented Kalman filter to estimate states and model uncertainties simultaneously by linear combinations of physics-motivated library functions. Using a sparsity promoting approach, a selection of those linear combinations is chosen and thus an interpretable model can be extracted. Results indicate a small estimation error compared to a traditional square-root unscented Kalman filter and exhibit the enhancement of physically meaningful models.  
\end{abstract}

\begin{keyword}
joint estimation, unscented transform, Kalman filter, sparsity, data-driven, compressed sensing 
\end{keyword}
\end{frontmatter}
\section{Introduction}
Estimating relevant quantities is an underlying key issue of many engineering problems, e.g. estimating a vehicle's position or velocity for tracking control in driver assistance systems. Since its introduction by \cite{Kalman.1960}, the Kalman filter has become one of the most prevalent estimators due to its optimality for linear systems. Naturally, numerous filter derivations for nonlinear dynamics were introduced thereafter, e.g. the extended Kalman filter (EKF), already used for the Apollo  missions to the moon, and the unscented Kalman filter (UKF) by \cite{Julier.1997}. 
However, an accurate model is crucial for the estimation performance regardless of which model-based filter is deployed. Recalling complex plants or systems, the process of modeling does not become difficult or time-consuming only due to parameter identification but due to complicated partial dynamics, often containing highly nonlinear parts and being challenging to capture.
Though many powerful filter techniques exist in control engineering, they strongly rely on the quality of the plant's model and will provide faulty estimates if it is inaccurate. 
Therefore, a filter that is able to cope with a low-quality model and its uncertainty while also returning a small estimation error and a comprehensive model would be desirable.

Regarding parameters uncertainty, concepts like joint and dual estimation are sufficiently known, e.g. \cite{Wan.1999}, but only few studies on filter approaches dealing with complex model uncertainties and their detection have been carried out so far.  Recently, only \cite{Khajenejad.2021}, \cite{BuissonFenet.2021} and \cite{Kullberg.2021b} investigated situations when just a partial model of the system is known. \cite{Khajenejad.2021} utilize a complex geometrical approach to estimate an upper and lower barrier for the approximated state to lie in between. The states and unknown partial dynamics are estimated very closely, but  the approach does not enable an interpretable representation of the unknown dynamics.
In \cite{BuissonFenet.2021}, a high gain observer is deployed to estimate states that are then utilized to infer a Gaussian process model of the plant. Subsequently, the model is used within the observer and vice versa, affecting and correlating with each other. Only \cite{Kullberg.2021b} strive for a similar strategy as we do: They propose linear combinations of radial basis functions (RBFs) with compact support to approximate the unknown partial dynamics. This is carried out by extending the state with the linear combinations' parameters to infer dominant RBFs. Established within an EKF, the authors focus on an efficient computation by altering the EKF formulas rather than identifying and interpreting the unknown partial dynamics while simultaneously estimating the states.

In this work, we start from a problem formulation similar to \cite{Khajenejad.2021}, \cite{BuissonFenet.2021} and \cite{Kullberg.2021b} and define a joint state, that additionally contains parameters of the linear combinations for the partial dynamics to be approximated. However, we use a square-root UKF (SQ-UKF) combined with a physics-motivated library for the linear combinations, simplifying the inference of physical coherences. Finally, arguing similar to \cite{Brunton.2016} that most systems in nature and technology can be characterized by a small amount of dynamical terms, we identify a method from compressed sensing (CS), see e.g. \cite{Carmi.2010}, for our field of research and adjust it to estimate the full state and an interpretable representation of the system's unknown partial dynamics. In particular, we deploy the idea of pseudo measurements within a SQ-UKF to enforce the linear combinations' parameters being sparse. In contrast to the method in \cite{Brunton.2016} and related approaches such as LASSO (\cite{Tibshirani.1996}), the model selection is carried out online while estimating the system's states.   
Therefore, our main contributions are the following:
\begin{itemize}
	\item Joint estimation of states and model uncertainties within a SQ-UKF, using physics-based and experts' knowledge,
	\item a sparse strategy to promote interpretable models by methods of CS, 
	\item demonstration and discussion on relevant application systems.
\end{itemize}  

The paper proceeds as follows: In Sec. \ref{sec:problem} the challenge of estimating states while only a partial model of the considered system being available is discussed. Hence, the strategy of a sparse, joint SQ-UKF, that estimates states and model uncertainties simultaneously, is outlined and clarified by notions of CS in Sec. \ref{sec:sparseUKF}. The proposed filter is then demonstrated for experiments in Sec. \ref{sec:Simulations}. A short discussion and outlook in Sec. \ref{sec:conclusion} concludes this work.

\textit{Notation:} $\bld{\tilde{\bullet}}$ denotes the joint state vector or relating variables, such as the covariance matrix. A subscript $\bullet_{k\vert k-1}$ represents the variable at time step $k$ with observations up to time step $k-1$. Bold variables refer to vectors or matrices. 

\section{Problem Formulation}\label{sec:problem}
For state estimation we wish to obtain the state $\bld{x}_k\in\mathbb{R}^{n_{\bld{x}}}$ at time step $k$ of a real system, whose dynamics are modeled by the following discrete system with control input $u_k\in\mathbb{R}$
\begin{align}\label{eq:model}
\begin{split}
	\bld{x}_{k+1}&=\bld{f}(\bld{x}_k,u_k)+\bld{w}_k,\\
	\bld{y}_k&= \bld{h}(\bld{x}_k,u_k)+\bld{v}_k.
\end{split}
\end{align}
The real system is assumed to be observable and can be observed by measurements $\bld{y}_k\in\mathbb{R}^m$.
Here, $\bld{f}$ and $\bld{h}$ denote the dynamical and observational models, respectively, with additive process and measurement noise $\bld{w}_k\sim\mathcal{N}(\bld{0},\bld{Q})$ and $\bld{v}_k\sim\mathcal{N}(\bld{0},\bld{R})$ assumed to be Gaussian distributed. 
However, if the underlying model \eqref{eq:model} lacks of relevant quantities, the estimator based upon that model will not be able to capture the non-measurable states closely or will diverge. 

Nonetheless, we could assume the unknown partial dynamics $\bld{g}(\bld{x},u)$ as a mapping that can be approximated by linear combinations of suitable functions $\Psi_i$, stored in a library $\bld{\Psi}$, and evaluated by parameters $\bld{\theta}\in\mathbb{R}^{n_{\bld{\theta}}}$. These functions contain prior knowledge, resulting from observational experience or experts' hypotheses, so that a close approximation by $\bld{g}(\bld{x},u)\approx\bld{\theta}^T\cdot\bld{\Psi}(\bld{x},u)$ can be expected. Presuming that the real system's order and its states can be defined, $\bld{\Psi}$'s minimal number of elements is $n_{\bld{x}}+2$, as it contains at least constants, the states and the control input. However, $\bld{\theta}$'s values are initially unknown and need to be calculated. Regarding parameter identification the joint estimation scheme is sufficiently known for estimating states and parameters simultaneously during the filter iterations, see e.g. \cite{Wan.1999} and \cite{vanderMerwe.2001}. Therefore, we extend the model \eqref{eq:model} to estimate unknown partial dynamics by defining the parameters $\bld{\theta}$ within a joint state vector $\tilde{\bld{x}}_k^T = \left(\bld{x}_k^T,\bld{\theta}_k^T\right)^T\in\mathbb{R}^{\tilde{n}}$ with $\tilde{n}=n_{\bld{x}}+n_{\bld{\theta}}$:
\begin{align}\label{eq:JEmodel}
	\begin{split}
	\tilde{\bld{x}}_{k+1}&=\begin{pmatrix}
	\bld{f}(\bld{x}_k,u_k,\bld{g}(\bld{x}_k,u_k))+\bld{w}_k^{\bld{x}}\\
	\bld{\theta}_k+\bld{w}_k^{\bld{\theta}}
	\end{pmatrix}\\
	&=\begin{pmatrix}
	\bld{f}(\bld{x}_k,u_k,\bld{\theta}_k^T\bld{\Psi}(\bld{x}_k,u_k))+\bld{w}_k^{\bld{x}}\\
	\bld{\theta}_k+\bld{w}_k^{\bld{\theta}}
	\end{pmatrix},\\
	\bld{y}_k&=\bld{h}(\bld{x}_k,u_k)+\bld{v}_k.	
	\end{split}
\end{align}  
Remark that $\bld{\theta}$'s evolution is modeled by stationary dynamics and additive Gaussian noise $\bld{w}_k^{\bld{\theta}}\sim\mathcal{N}(\bld{0},\bld{Q}_{\bld{\theta}})$, whereas it holds $\bld{w}_k^{\bld{x}}\sim\mathcal{N}(\bld{0},\bld{Q}_{\bld{x}})$. Hence, the process covariance matrix of the joint state is a block matrix $\bld{\tilde{Q}}= \text{blkdiag}(\bld{Q}_{\bld{x}},\bld{Q}_{\bld{\theta}})$ with zeros elsewhere. Note within the scope of this paper the unknown partial dynamics are considered as one dimensional, namely $g(\bld{x},u)$, and $\bld{h}$ is linear. To receive a small estimation error $\bld{e}_k =\bld{\hat{\tilde{x}}}_k-\bld{\tilde{x}}_k$, any Kalman filter relies on  the minimum-mean-squared-error to estimate $\bld{\hat{\tilde{x}}}_k$ as stated e.g. in \cite{Gibbs.2011}:
\begin{equation}\label{eq:filterProblem}
\argmin_{\bld{\hat{\tilde{x}}}} \frac{1}{2}\mathbb{E}[(\bld{\hat{\tilde{x}}}-\bld{\tilde{x}})^T(\bld{\hat{\tilde{x}}}-\bld{\tilde{x}})].
\end{equation}
Thus, the lift towards a higher dimensional state $\bld{\tilde{x}}$, including the unknown partial dynamics identifying parameters $\bld{\theta}$, affects the optimization problem \eqref{eq:filterProblem} by the cost of increased model complexity and computational burden. 

\section{Joint Estimation by sparsity}\label{sec:sparseUKF}
After the model's reformulation to estimate states and unknown partial dynamics simultaneously, this section copes with the model's application within an SQ-UKF and presents a sparse solution as remedy towards the computational burden and the degrees of freedom due to the high dimensional state. 

\subsection{Square-root unscented Kalman filter}
The unscented Kalman filter (UKF), established by \cite{Julier.1997, Julier.2002}, is a powerful nonlinear filter design method. In contrast to the extended Kalman filter (EKF), which linearizes nonlinear dynamics along a single sample point by the first order Tailor series, the UKF seeks to estimate the moments of the nonlinear probability distribution by multiple sample points. Therefore, it is independent of the analytical differentiation calculations the EKF requires. This enables a higher estimation accuracy when systems with strong nonlinearities are considered. However, the calculation of the covariance matrix in every filter iteration for $2\tilde{n}+1$ samples, called sigma points, is computational expensive. But, as presented in \cite{vanderMerwe.2001,vanderMerwe.2004}, the UKF's efficiency and its numerical stability can be improved by using the square-root covariance matrix $\bld{S}_k$ instead of $\bld{P}_k=\bld{S}_k\bld{S}_k^T$, leading to the square-root UKF (SQ-UKF) algorithm. 

In Algo. \ref{algo:UKF} the SQ-UKF's procedure is outlined. 
To ensure a symmetric distribution during the unscented transformation, the UKF's parameters $\lambda=\alpha^2(\tilde{n}+\kappa)-\tilde{n}$ and $\eta=\sqrt{\tilde{n}+\lambda}$ define the weights $\bld{W}^{(c)}$ and $\bld{W}^{(m)}$: 
\begin{align}\label{eq:UKFparameters}
\begin{split}
	W_0^{(m)}&=\frac{\lambda}{\lambda+\tilde{n}},\;
	W_0^{(c)}=\frac{\lambda}{\lambda+\tilde{n}}+1-\alpha^2+\beta,\\
	W_i^{(m)}&=W_i^{(c)}=\frac{1}{2(\tilde{n}+\kappa)}\quad\text{for}\,i=1,\dots,2\tilde{n}.
\end{split}	
\end{align}
The parameters $\alpha, \beta$ and $\kappa$ need to be chosen beforehand, since they reflect how the nonlinear distribution is covered. In \cite{Nielsen.2021} it has been shown that the choice of these is problem-dependent and has a significant influence on the filter performance. However, for the scope of this paper, we choose $\alpha=10^{-3}, \beta=2$ and $\kappa=0$.

\begin{algorithm}
	\caption{SQ-UKF}\label{algo:UKF}
	\begin{algorithmic}[1]
		\STATE \textbf{Initialize:} $\alpha, \beta, \kappa,\bld{\hat{\tilde{x}}}_0=\mathbb{E}[\bld{\tilde{x}}_0],$\\
		\qquad\quad\quad\,	 $\bld{S}_0 = \text{chol}(\mathbb{E}[(\bld{\tilde{x}}_0-\bld{\hat{\tilde{x}}}_0)^T(\bld{\tilde{x}}_0-\bld{\hat{\tilde{x}}}_0)])$
		\item[]
		\STATE \textbf{for} {$k\in \{1,\dots,\infty\}$}\\
			\quad Calculation of sigma points:
		\STATE $\quad\mathcal{X}_{k-1}=[\bld{\tilde{x}}_{k-1}\quad\bld{\tilde{x}}_{k-1}+\eta\bld{S}_{k-1}\quad\bld{\tilde{x}}_{k-1}-\eta\bld{S}_{k-1}]$\\
		\quad Prediction:
		\STATE $\quad\mathcal{X}_{k\vert k-1}=\bld{f}(\mathcal{X}_{k-1},u_{k-1})$
		\STATE $\quad\bld{\hat{\tilde{x}}}_k^- =\sum_{i=0}^{2\tilde{n}} W_i^{(m)}\mathcal{X}_{i,k\vert k-1}$
		\STATE $\quad\bld{S}_k^-=\text{qr}\left(\left[\sqrt{W_1^{(c)}}(\mathcal{X}_{1:2\tilde{n},k\vert k-1}-\bld{\hat{\tilde{x}}}_k^-)\quad\sqrt{\bld{\tilde{Q}}}\right]\right)$
		\STATE $\quad\bld{S}_k^-=\text{cholupdate}(\bld{S}_k^-,\mathcal{X}_{0,k}-\bld{\hat{\tilde{x}}}_k^-,W_0^{(c)})$
		\STATE $\quad\mathcal{Y}_{k\vert k-1}=\bld{h}(\mathcal{X}_{k\vert k-1},u_{k-1})$
		\STATE $\quad\bld{\hat{{y}}}_k^- =\sum_{i=0}^{2\tilde{n}}W_i^{(m)}\mathcal{Y}_{i,k\vert k-1}$\\
		\quad Correction:
		\STATE $\quad\bld{S}_{y_k}=\text{qr}\left(\left[\sqrt{W_1^{(c)}}(\mathcal{Y}_{1:2\tilde{n},k\vert k-1}-\bld{\hat{\tilde{x}}}_k^-)\quad\sqrt{\bld{R}}\right]\right)$
		\STATE $\quad\bld{S}_{y_k}=\text{cholupdate}(\bld{S}_{y_k},\mathcal{Y}_{0,k}-\bld{\hat{{y}}}_k^-,W_0^{(c)})$
		\STATE $\quad\bld{P}_{xy}=\sum_{i=0}^{2\tilde{n}}W_i^{(c)}[\mathcal{X}_{i,k\vert k-1}-\bld{\hat{\tilde{x}}}_k^-][\mathcal{Y}_{i,k\vert k-1}-\bld{\hat{{y}}}_k^-]^T$
		\STATE $\quad\bld{K}_k =\bld{P}_{xy}\bld{P}_{yy}^{-1}$
		\STATE $\quad\bld{\hat{\tilde{x}}}_k=\bld{\hat{\tilde{x}}}_k^- +\bld{K}_k(\bld{y}_k-\bld{\hat{{y}}}_k^-)$
		\STATE $\quad\bld{U}=\bld{K}_k\bld{S}_{y_k}$
		\STATE $\quad\bld{S}_k=\text{cholupdate}(\bld{S}_k^-, \bld{U}, -1)$
		\STATE \textbf{end}
	\end{algorithmic}
\end{algorithm}

If the SQ-UKF is utilized in its original form (Algo. \ref{algo:UKF}) without any modifications for the additional degrees of freedom resulting from $\bld{\theta}$, it has been observed that the filter tends to become locally unstable or divergent, which may lead to faulty state estimates. Depending on the number of design degrees $n_{\bld{\theta}}$, the calculation of $\bld{\theta}$ comes close to an underdetermined optimization problem. Hence, a strategy to choose only a selection of $\bld{\theta}$ while retaining the design degrees is required. Therefore, notions from the field of compressed sensing are adopted.
  
\subsection{Compressed sensing}\label{sec:cs}
In the domain signal processing methods of compressed sensing cope with the reconstruction of signals $\bld{x} =\bld{\Omega}\bld{s}$ with $\bld{\Omega}\in\mathbb{R}^{n\times n}$ a suitable basis matrix and $\bld{s}\in\mathbb{R}^n$ the sparse representation of the signal $\bld{x}\in\mathbb{R}^n$. Based on early works of \cite{Donoho.2006} and \cite{Candes.2006}, the linear underdetermined problem with the measurement $\bld{y}=\bld{H}\bld{x}$ and its measurement matrix $\bld{H}$ 
\begin{equation}\label{eq:l0}
	\argmin_{\bld{s}} \vert\vert \bld{H}\bld{\Omega}\bld{s}-\bld{y}\vert\vert_2+\tilde{\lambda}\vert\vert\bld{s}\vert\vert_0,
\end{equation}  
which seeks for a sparse solution $\bld{s}$ and is non-convex due to the $\ell_0$-norm leading to a high combinatorial effort, can be transformed into 
\begin{equation}\label{eq:l1}
	\argmin_{\bld{s}} \vert\vert \bld{H}\bld{\Omega}\bld{s}-\bld{y}\vert\vert_2+\tilde{\lambda}\vert\vert\bld{s}\vert\vert_1.
\end{equation}
Here, $0<\tilde{\lambda}\ll1$ denotes the sparsity balancing factor. The minimization problem \eqref{eq:l1} is now convex and provides a solution if sufficiently many measurements $\bld{y}$ are given and the rows of $\bld{H}$ are independent of $\bld{\Omega}$'s columns, for details see \cite{Carmi.2010,Brunton.2019}. Further, the problem can be reformulated into a constrained optimization problem
\begin{align}\label{eq:l1-relax}
\begin{split}
	&\argmin_{\bld{s}} \vert\vert \bld{H}\bld{\Omega}\bld{s}-\bld{y}\vert\vert_2,\\
s.t.\,&\vert\vert\bld{s}\vert\vert_1\leq\epsilon,
\end{split}
\end{align}
with $0<\epsilon\ll1$.

\subsection{Promoting sparsity by pseudo measurements}

Considering the degrees of freedom introduced by the parameters $\bld{\theta}$, an accurate estimation by the filter problem \eqref{eq:filterProblem} can not be guaranteed any more as discussed in the previous sections.
However, using the same argument as in \cite{Brunton.2016}, most phenomena in nature or technology can be represented rather by a small amount of dominating terms. Therefore, only a selection of $\bld{\Psi}$ is needed to capture the system's behavior. To enforce this assumption, the filter problem \eqref{eq:filterProblem} is reformulated by adding a term that encourages a sparse vector $\bld{\theta}$. Adopting the notions \eqref{eq:l0} and \eqref{eq:l1} from Sec. \ref{sec:cs} to efficiently propagate a sparse $\bld{\theta}$ within the iterative filter Algo. \ref{algo:UKF}, the minimization problem can be formulated by
\begin{equation}\label{eq:filterProblemSparse}
\argmin_{\bld{\hat{\tilde{x}}}} \frac{1}{2}\mathbb{E}[(\bld{\hat{\tilde{x}}}-\bld{\tilde{x}})^T(\bld{\hat{\tilde{x}}}-\bld{\tilde{x}})] + \tilde{\lambda} \vert\vert\bld{\hat{\tilde{x}}}_{n+1:\tilde{n}}\vert\vert_1
\end{equation} 
with $\tilde{\lambda}\in\mathbb{R}$ as the sparsity promoting term. 
However, how to incorporate this approach within the iterative structure of the SQ-UKF is not straight-forward. Recalling the previous section and adjusting \eqref{eq:l1-relax} towards the problem \eqref{eq:filterProblemSparse} yields a sparse mean-squared-error for joint estimation 
\begin{align}\label{eq:CS}
\begin{split}
&\argmin_{\bld{\hat{\tilde{x}}}} \frac{1}{2}\mathbb{E}[(\bld{\tilde{x}}-\bld{\hat{\tilde{x}}})^T(\bld{\tilde{x}}-\bld{\hat{\tilde{x}}})] \\
s.t. \,&\vert\vert\bld{\hat{\tilde{x}}}_{n+1:\tilde{n}}\vert\vert_1 \leq \epsilon.
\end{split}
\end{align}
By interpreting the soft constraint as a fictitious, additional measurement $0= \vert\vert\bld{\hat{\tilde{x}}}_{n+1:\tilde{n}}\vert\vert_1-\epsilon$ it can be utilized during estimation schemes, see e.g. \cite{Julier.2007}, \cite{Carmi.2010}, \cite{Hage.2020b}. Different approaches, e.g. as shown in \cite{Julier.2007}, exist how to incorporate the pseudo measurement within a filter iteration. For this paper, we define a pseudo measurement function $\tilde{h}(\bld{\tilde{x}})$ by
\begin{equation}\label{eq:pseudoh}
\tilde{h}(\bld{\tilde{x}}) = \vert\vert\bld{\hat{\tilde{x}}}_{n+1:\tilde{n}}\vert\vert_1-\epsilon, 
\end{equation}
where $\epsilon\sim\mathcal{N}(0,R_{pm})$ now represents the fictitious, additive measurement noise and controls the tightness of the constraint. Since the SQ-UKF is based on the unscented transformation, the nonlinear observational model \eqref{eq:pseudoh} can be easily applied without any further adaption in contrast to its use within an EKF, e.g. \cite{Julier.2007}.   

In Algo. \ref{algo:sparseUKF} the procedure of the proposed sparsity promoting, joint SQ-UKF (J-SQ-UKF) is depicted. Within every filter iteration $k$ additional updates for the estimated state $\bld{\hat{\tilde{x}}}_k$ and its square-root covariance $\bld{S}_k$ might be enclosed depending on the sparsity of $\bld{\theta}_k$. If the number $n_{\bld{\theta},act}$ of non-sparse elements, that are greater than the sparsity barrier $\tilde{\lambda}$, is exceeded (line 18), a filter update is calculated again but now using the pseudo measurement \eqref{eq:pseudoh} as fictitious observational model instead of the actual observational model $\bld{h}$ (line 19). This step is followed by additional checks and updates. The sparsity promoting loop terminates either when the maximal iteration of pseudo measurements $j=N$ is reached or the number of allowed non-sparse elements is less or equal $n_{\bld{\theta},act}$. In contrast to Sec. \ref{sec:cs} online estimation is possible because of the evolving covariance that contains information about past measurements.
Since with every filter iteration $k$ the dominant parameters of $\bld{\theta}$ could completely change due to the sparsity promoting loop and the degrees of freedom at the beginning of every iteration $k$, the stability of the filter algorithm may be easily affected as well as the model's interpretability might get lost. Hence, to avoid local instabilities or divergent behavior, a soft switching method for $\bld{\theta}$ is enforced afterwards (line 23) by balancing the state estimate before and after the pseudo measurement correction. 
This guarantees a more continuous estimation of the dominant parameters, resulting in a more continuous state estimation itself. 
Different methods of soft switching could be applied, e.g. batch updates, but herein a simple weighted approach is employed by a factor $0<\gamma<1$. It is recommended to enforce a higher emphasis on the newly corrected estimate after the pseudo measurement loop. 

\begin{algorithm}
	\caption{J-SQ-UKF}\label{algo:sparseUKF}
	\begin{algorithmic}[1]
		\setcounter{ALC@line}{14}
		\STATE \quad$\vdots$
		\STATE $\quad\bld{S}_k=\text{cholupdate}(\bld{S}_k^-, \bld{U}, -1)$\\
		\item[]
		\quad Sparsity promoting update:
		\STATE \quad\textbf{Initialize:} ${\tilde{h}}, N, n_{\bld{\theta},act},\gamma,j = 1$\\ 
		$\quad\quad\quad\quad\quad\quad\bld{S}_{pm,0}= \bld{S}_k, \bld{\hat{\tilde{x}}}_{pm,0}=\bld{\hat{\tilde{x}}}_k$
		\item[]
		\STATE $\quad$\textbf{while} {$\#\{\theta_j\vert\theta_j>\tilde{\lambda}\}>n_{\bld{\theta},act}\;\textbf{and}\; j<N$}
		\STATE $\quad\quad \bld{\hat{\tilde{x}}}_{pm,j},\bld{S}_{pm,j}\leftarrow\text{SQ-UKF}\; \text{with}$\\
	$\quad\quad\quad\quad\quad\quad\quad\quad\quad(\bld{\hat{\tilde{x}}}_{pm,j-1},\bld{S}_{pm,j-1},\bld{f},\tilde{h},\bld{\tilde{Q}},R_{pm})$
		\STATE $\quad\quad j = j+1$\\
		\STATE $\quad$\textbf{end}
		\item[]
		\STATE $\quad\bld{S}_{k,final}=\bld{S}_{pm,j}$
		\STATE $\quad[\bld{\hat{\tilde{x}}}_{k,final}]_{1:n_{\bld{x}}}=[\bld{\hat{\tilde{x}}}_k]_{1:n_{\bld{x}}},$\\ $\quad[\bld{\hat{\tilde{x}}}_{k,final}]_{n_{\bld{x}}+1:\tilde{n}}=(1-\gamma)[\bld{\hat{\tilde{x}}}_{pm,j}]_{n_{\bld{x}}+1:\tilde{n}}+\gamma[\bld{\hat{\tilde{x}}}_k]_{n_{\bld{x}}+1:\tilde{n}}$
		\STATE \textbf{end}
	\end{algorithmic}
\end{algorithm}

\subsection{Curse of dimension}
As generally known, any Kalman filter's performance depends strongly on the choice of its covariance matrices. While the measurement covariance $\bld{R}\in\mathbb{R}^{m\times m}$ often is easier to determine by taking a closer look at the plant's measurements, the initial square-root covariance $\bld{S}_0$ and the process covariance $\bld{\tilde{Q}}\in\mathbb{R}^{\tilde{n}\times\tilde{n}}$ remain a major challenge in filter design, see e.g. \cite{Chen.2021, Nielsen.2021}. When promoting sparsity within the SQ-UKF even an additional covariance $R_{pm}\in\mathbb{R}$ comes along which determines the pseudo measurements' performance. With increasing numbers of states $n_{\bld{x}}$ and possibly uncertain partial dynamics, expressed by $n_{\bld{\theta}}$ numerous parameters $\bld{\theta}$, not only the curse of dimension but also the difficulty to initialize the performance parameters become obvious. Common approaches in filter design include manual tuning, grid or random search, which are unsatisfying in terms of effort and benefit while possibly resulting in local minima. Lately, Bayesian optimization becomes the state-of-the-art method for hyperparameter optimization and could be applied herein when offline estimation is pursued. For the scope of this paper, we focus on online estimation and therefore set the parameters as $\bld{P}_0=\bld{\tilde{Q}}=\text{blkdiag}(10^{-6}\bld{I}_{n_{\bld{x}}},10^{-4}\bld{I}_{n_{\bld{\theta}}}), \bld{R}=10^{-4}\bld{I}_m$, and $R_{pm}=1$.

\section{Results}\label{sec:Simulations}
The proposed filter algorithm is now demonstrated for two application examples. Compared to a traditional SQ-UKF, the advantages of the sparse, joint SQ-UKF filter become very clear. Within each application example both filters start with a corrupted initial state $\bld{\hat{\tilde{x}}}_0\neq \bld{\tilde{x}}_0$. For the J-SQ-UKF the parameters $\bld{\theta}_0$ are initialized equally with small values. Within all experiments discussed in this paper it holds $n_{\bld{\theta},act}=3$ and $\tilde{\lambda}=0.1$.  

\subsection{Duffing oscillator}
The Duffing oscillator is a common example for nonlinear dynamical systems. Its dynamics are described in the following with parameters $\bld{p}=(-1,3,0.1)^T$:
\begin{align}\label{eq:Duffing}
\begin{split}
\bld{\dot{x}}&=\begin{pmatrix}
x_2\\ -p_3x_2-p_1x_1-p_2x_1^3+u
\end{pmatrix},\\
y&=x_1.
\end{split}
\end{align} 
Here, we assume the nonlinear term $g(\bld{x},u)=p_2x_1^3$ as unknown and formulate the joint model, lacking of the mentioned term, analogously to the description in Sec. \ref{sec:problem}. Formulating \eqref{eq:Duffing} as a discrete model and using the explicit Euler scheme and the library $\bld{\Psi}_1=(1,x_1,x_2,x_2^2,\sin(x_2),x_1^3,x_1x_2,\cos(x_1),u)^T$ to evaluate the model, we can deploy it within the J-SQ-UKF. 
\begin{figure}[h]
	\centering
	\includegraphics[width=\columnwidth]{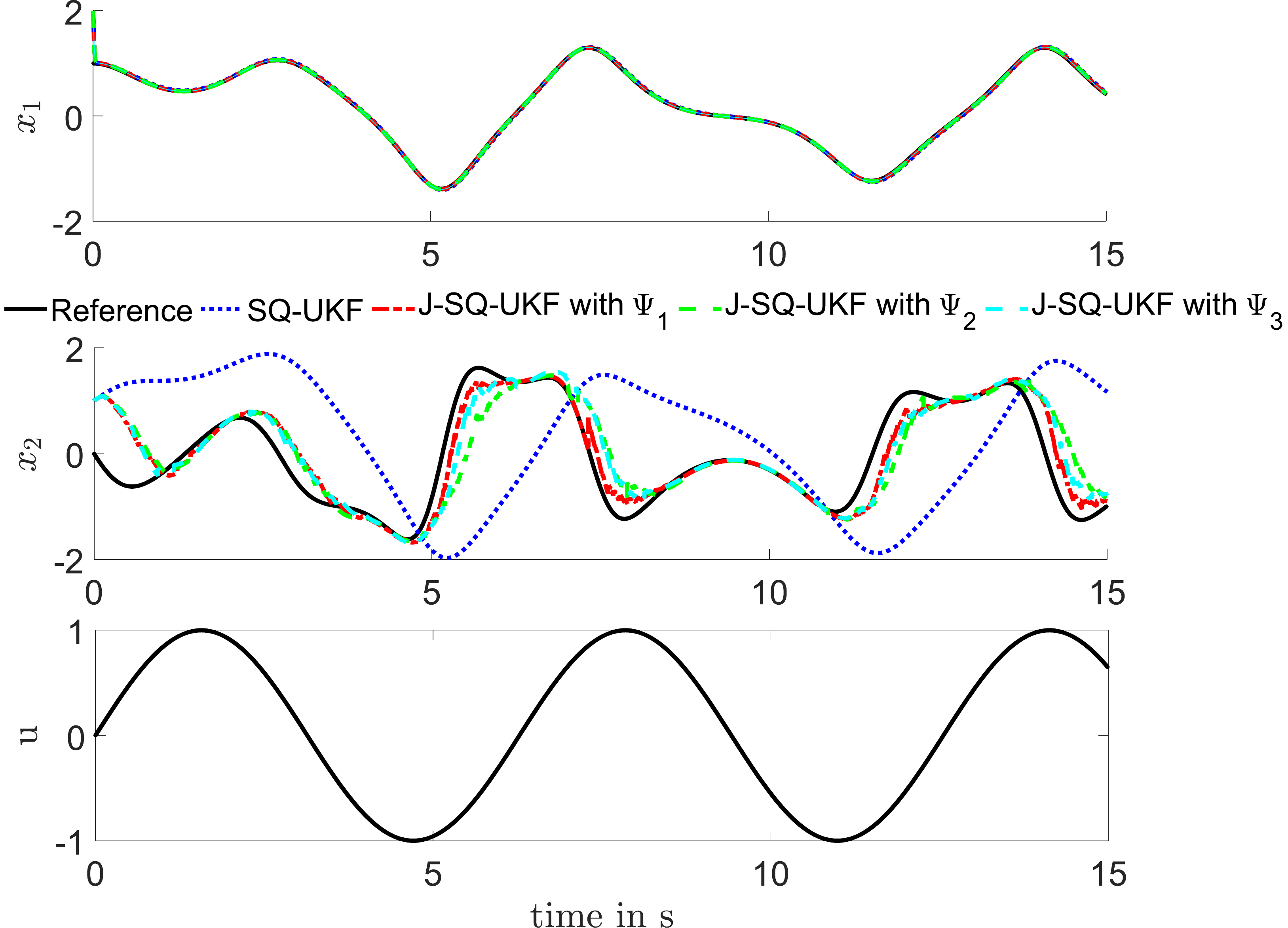}
	\caption{Estimation of the Duffing oscillator's states with sine excitation comparing various filter performances}\label{fig:duffingJE}
\end{figure}

\begin{figure}
	\centering
	\subfloat[Dominant term $\theta_6$: $\Psi_6(\bld{x},u)=x_1^3$]{
	\includegraphics[width=0.9\columnwidth]{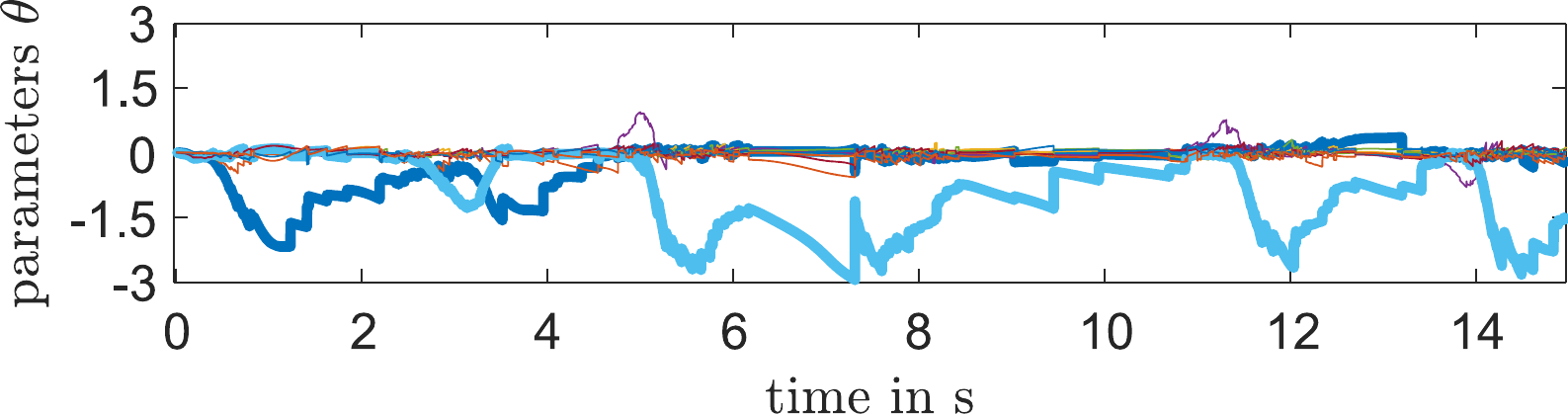}}\\
	\subfloat[Dominant term $\theta_2$: $\Psi_2(\bld{x},u)=x_1$]{
		\includegraphics[width=0.9\columnwidth]{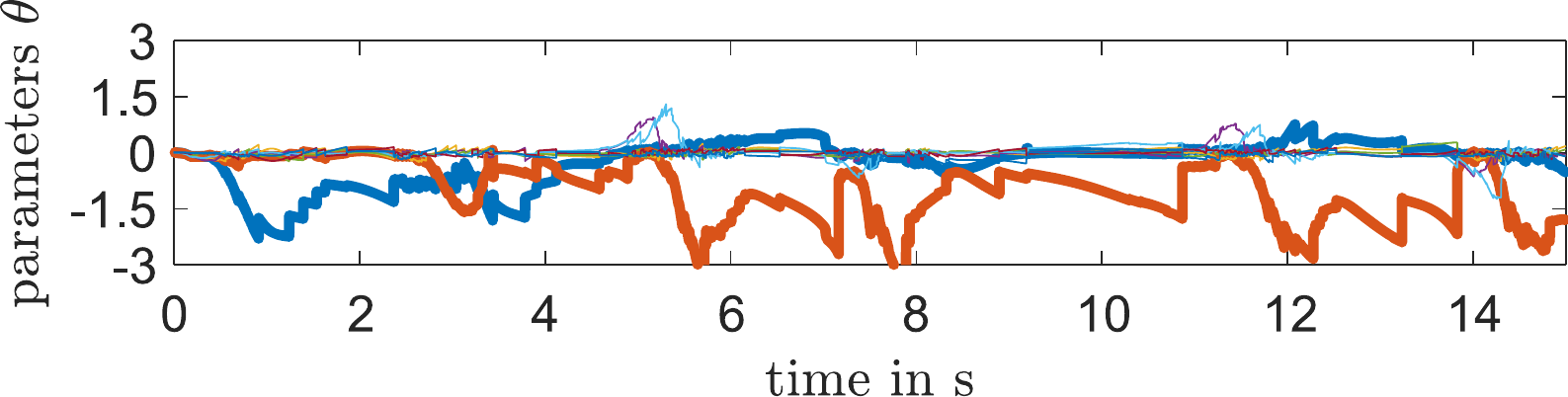}}\\
\subfloat[Dominant term $\theta_6$: $\Psi_6(\bld{x},u)=x_1^2$]{
	\includegraphics[width=0.9\columnwidth]{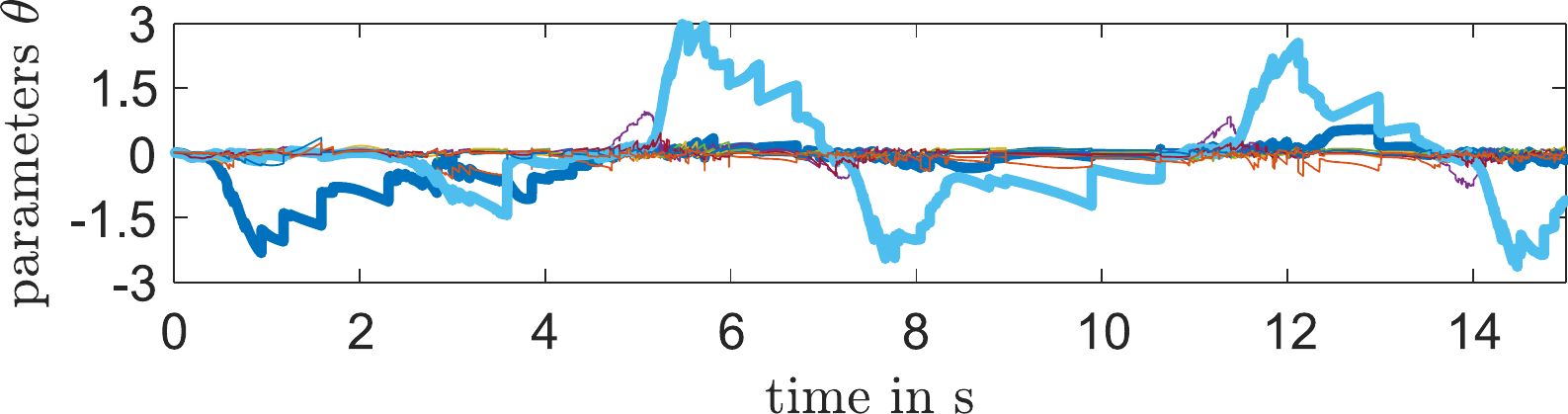}}
	\caption{Course of $\bld{\theta}$ corresponding to different libraries $\bld{\Psi}_1,\bld{\Psi}_2$ and $\bld{\Psi}_3$ from top to bottom: After a transient phase the dominant $\theta_i$ indicates which term is best for $g$ to be approximated.}\label{fig:duffingJEtheta}
\end{figure}

As a result, we see through the red dashed line in Fig. \ref{fig:duffingJE} that the J-SQ-UKF is able to cope with the faulty model and to deliver a correct estimate. This is due to the selection of the true dynamics, which is illustrated in the top plot in Fig. \ref{fig:duffingJEtheta} by $\bld{\theta}$'s magnitude over time. Here, the line, that represents $\theta_6$ and therefore indicates that $\Psi_6(\bld{x},u)=x_1^3$ is mainly present, is of the highest magnitude. Further, it is able to deal with incorrect initial states and indicates a fast transient behavior towards the true states. Conversely, the traditional SQ-UKF is not able to handle the model uncertainty and even provides incorrect estimates regarding the initial state as depicted in Fig. \ref{fig:duffingJE} by the blue dotted line.

When utilizing a library without the true term $g$, e.g. $ \bld{\Psi}_2=(1,x_1,x_2,x_2^2,\sin(x_2),x_1x_2,\cos(x_1),u)^T$ or $\bld{\Psi}_3=(1,x_1,x_2,x_2^2,\sin(x_2),x_1^2,x_1x_2,\cos(x_1),u)^T$, the J-SQ-UKF is yet able to estimate the true dynamics. As illustrated in Fig. \ref{fig:duffingJE} and \ref{fig:duffingJEtheta}, the filter clearly extracts other physical terms that are close to the true missing term $g$ and still delivers an accurate state estimate, though a small degradation compared to the J-SQ-UKF with $\bld{\Psi}_1$ can be observed. Therefore, the proposed method appears to be confident in detecting either the correct partial dynamics or finding alternatives to interpret the system's behavior, namely $\Psi_2(\bld{x},u)=x_1$ or $\Psi_6(\bld{x},u)=x_1^2$. These interpretations can then be utilized for further model usage or refinement.

\subsection{Golf robot}
\begin{wrapfigure}[]{r}{0.25\columnwidth}
	\centering
	\includegraphics[width=0.25\columnwidth]{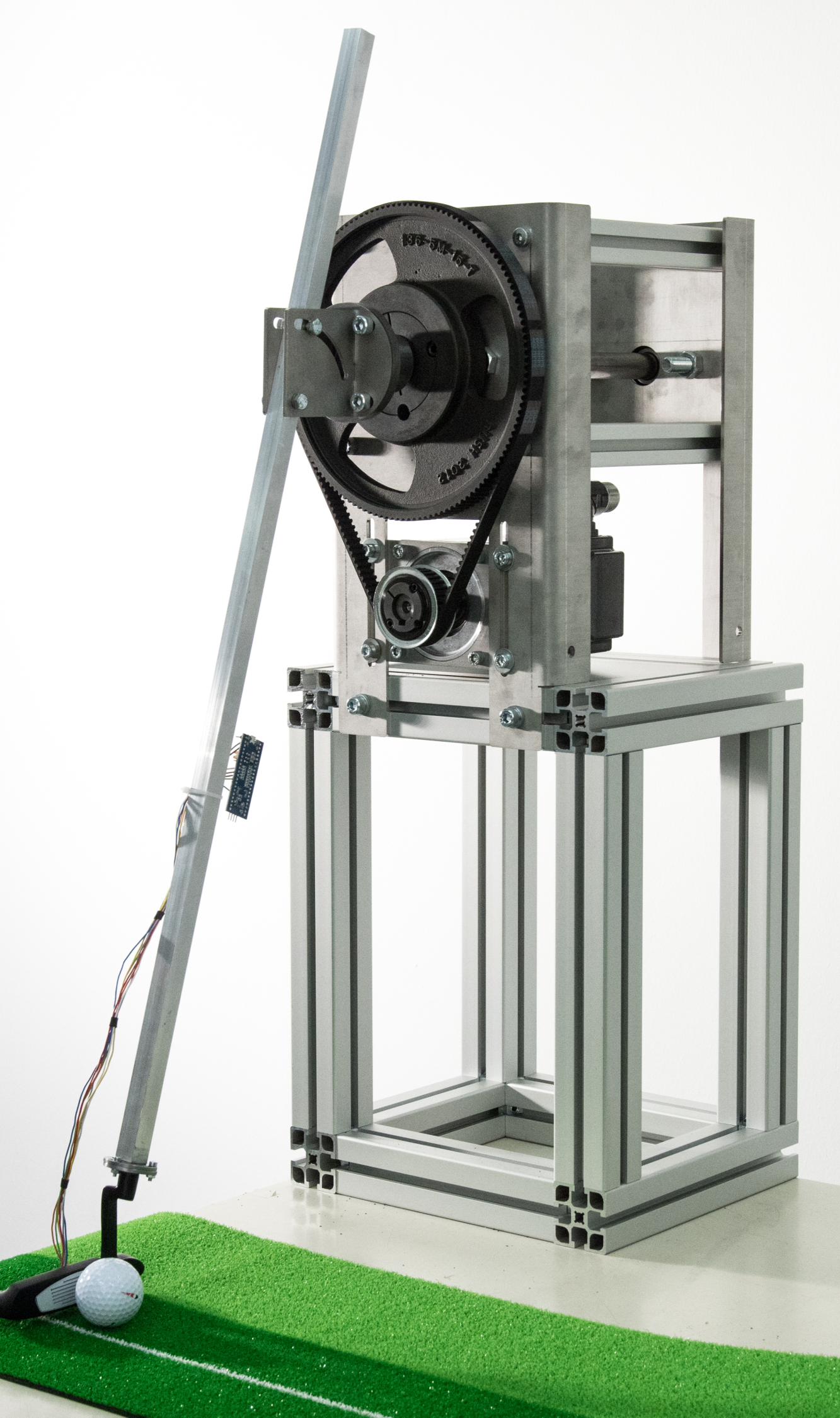}\caption{Golf robot}\label{fig:golfrobot}
\end{wrapfigure}

The golf robot, shown in Fig. \ref{fig:golfrobot}, is a test bench at our laboratory to investigate the capability of machine learning methods on real world problems. After extensive modeling effort the robot's dynamics are described by 
\begin{align}\label{eq:golfrobot}
\begin{split}
\bld{\dot{x}}
&= \begin{pmatrix}x_2\\
J^{-1}\left(-m \tilde{g} a\sin(x_1)- M_F+4u\right)
\end{pmatrix},\\
M_{F} &= dx_2+2r\mu \,\text{arctan}(10^3 x_2)\pi^{-1} \left| m x_2^2a + m \tilde{g}\cos\left(x_1\right)\right|,\\
y &= x_1,
\end{split}
\end{align}
with $\tilde{g}$ as gravity constant and the same parameters $\bld{p}=(m, a, d, J, r, \mu)^T$ utilized as in \cite{Gotte.2022, Schon.2022}. Here, the joint model \eqref{eq:JEmodel} is the model that completely lacks of the term $M_F$. Since the modeling effort of \eqref{eq:golfrobot} lies mainly in modeling the friction terms, this assumption is reasonable and close to real world situations. Further, the library is defined as $\bld{\Psi}(\bld{x},u)=(1,x_1,x_2,x_2^2,x_1^3,\sin(x_2),\cos(x_1),u)^T$ and the model is again discretized and evaluated with the explicit Euler scheme. 

\begin{figure}[h]
	\centering
	\includegraphics[width=\columnwidth]{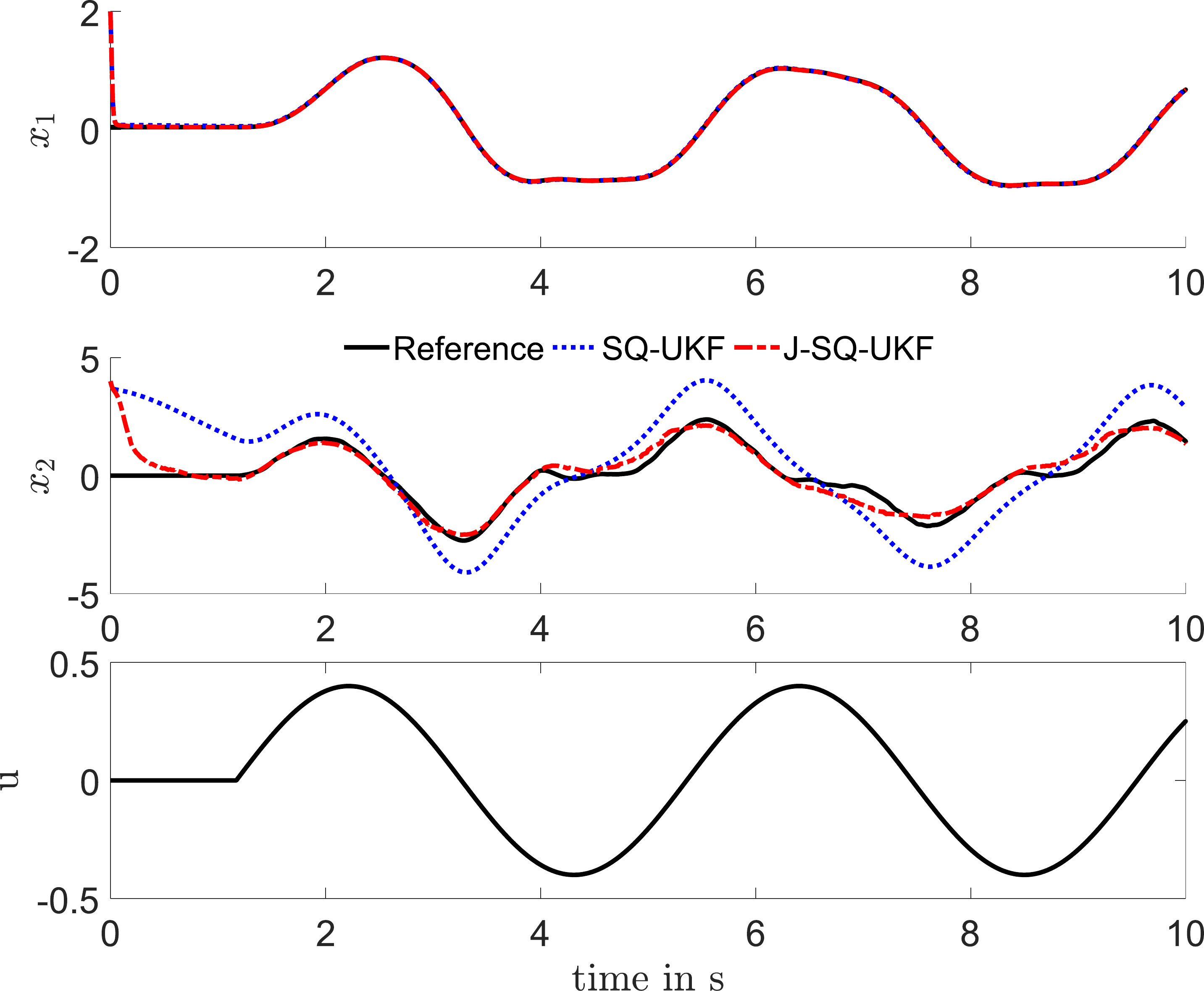}
	\caption{Estimation of the golf robot's states with sine excitation comparing various filter performances}\label{fig:golfstates}
\end{figure}

\begin{figure}[h]\centering
\includegraphics[width=\columnwidth]{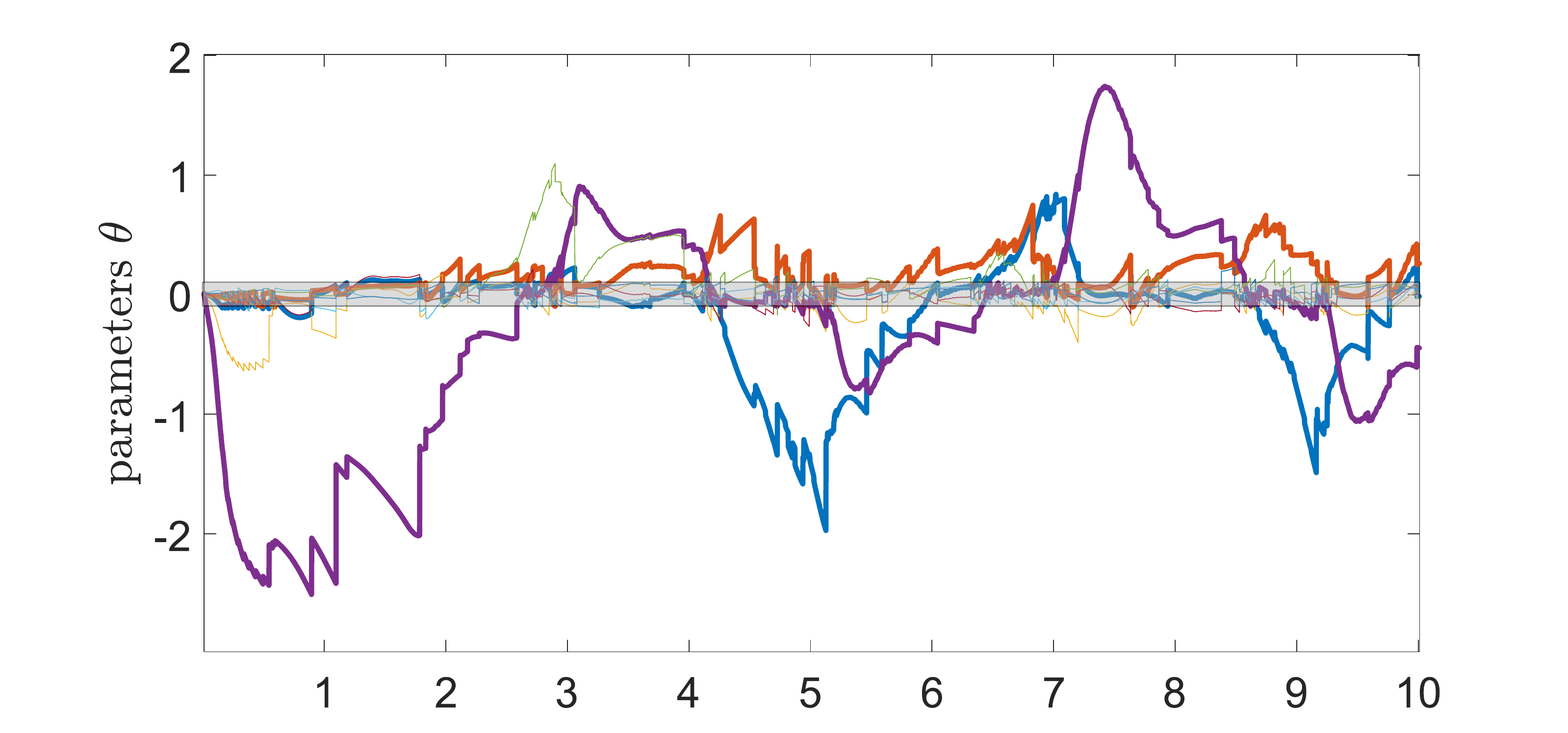}\caption{Course of $\bld{\theta}$ corresponding to library $\bld{\Psi}$: Dominant $\theta_i$ indicate that $\Psi_1(\bld{x},u)=1, \Psi_2(\bld{x},u)=x_1$ and $\Psi_4(\bld{x},u)=x_2^2$ are the best terms to approximate $g$. The barrier due to $\tilde{\lambda}=0.1$ is illustrated by the transparent gray rectangle.}\label{fig:golftheta}
\end{figure}

As illustrated in Fig. \ref{fig:golfstates} the J-SQ-UKF provides a close estimation of the robot's states and even shows a faster transient behavior than the traditional filter that delivers faulty angular velocities. The corresponding dominant parameters $\bld{\theta}$ are depicted in Fig. \ref{fig:golftheta} and reveal that more than one term of the library is relevant to approximate $g$, in fact these are $\Psi_1, \Psi_2$ and $\Psi_4$. Further, it is observed that these terms seem to vary in time which is reasonable since friction forces are dependent on velocity and are often highly nonlinear. It is also noteworthy that the dominant terms are close to the ones which have been identified so far by extensive modeling effort in model \eqref{eq:golfrobot} or correlate to those by factor, e.g. $x_1\approx\cos(x_1)$ when $x_1$ is small. However, extracting a more accurate representation for those dynamics requires a follow-up step like a modal transformation when there is more than one identified dominant term $\Psi_i$. 


\section{Discussion and outlook}\label{sec:conclusion}
The proposed method has shown for several application examples its power to jointly estimate states and model uncertainties. We could observe two cases: Either the library $\bld{\Psi}$ contains the unknown partial dynamics $g$ and the algorithm extracts the correct term or close approximations of it, or it provides a suggestion for $g$ based on its library terms.
Further, it clearly outperforms the traditional filter when only an incomplete model of the system is available. However, the interpretability has not been automated so far and needs the user to interpret and extract the dominant $\theta_i$ and its corresponding terms by sight. 
Additionally, parameters like the number of allowed sparse elements $n_{\bld{\theta},act}$, the sparsity barrier $\tilde{\lambda}$ and the terms of $\bld{\Psi}$ currently have to be set by the user. Correlations between those parameters are to be expected. Moreover, it has not yet been investigated if there exists a dependency of $n_{\bld{\theta}}$ and $n_{\bld{x}}$, limiting possibly the number of $\bld{\Psi}$'s elements. Nonetheless, the sparse, joint SQ-UKF allows a deeper insight into the system's dynamics, while simultaneously providing sufficient estimates. This insight could be then concretized offline by a modal transformation method, e.g. proper orthogonal decomposition, to automate the model interpretations and utilize an updated model within the filter. However, experiments have shown to be real-time capable and encourage the usage within online estimation. 
Therefore, future research focuses on the relation of $n_{\bld{\theta}}$ and $n_{\bld{x}}$ and on a higher dimensional $\bld{g}$.

\section*{Acknowledgment}

This work was developed in the junior research group DART (Daten\-ge\-trie\-be\-ne Methoden in der Regelungstechnik), University Paderborn, and funded by the Federal Ministry of Education and Research of Germany (BMBF - Bundesministerium für Bildung und Forschung) under the funding code 01IS20052. The responsibility for the content of this publication lies with the authors.

\end{document}